\documentstyle[graphicx]{mn}
\newcommand{\ls}
 {\mathrel{\hbox{\rlap{\hbox{\lower4pt\hbox{$\sim$}}}\hbox{$<$}}}}
\newcommand{\gs}
 {\mathrel{\hbox{\rlap{\hbox{\lower4pt\hbox{$\sim$}}}\hbox{$>$}}}}

\newcommand{\et}{et al.\ }

\newcommand{\asca}{{\it ASCA}}
\newcommand{\xte}{{\it RXTE}}
\newcommand{\sax}{{\it BeppoSAX}}
\newcommand{\xmm}{{\it XMM-Newton}}

\def\la{\mathrel{\hbox{\rlap{\hbox{\lower4pt\hbox{$\sim$}}}{\raise2pt\hbox{$
<$}}
}}}
\def\ga{\mathrel{\hbox{\rlap{\hbox{\lower4pt\hbox{$\sim$}}}{\raise2pt\hbox{$
>$}}
}}}

\date{May 2002, submitted to MNRAS}

\title{Extreme X-ray variability in the luminous quasar PDS 456}

\author[Reeves {\it et al.}]
{J.N. Reeves$^1$, G. Wynn$^{1}$, P.T. O'Brien$^1$, K.A. Pounds$^{1}$\\
$^1$Department of Physics and Astronomy, University of Leicester,
University Road, Leicester LE1 7RH; U.K.}

\begin{document}
\maketitle
\begin{abstract}

We present evidence from \sax\ and \xmm\ of extreme X-ray
variability in the high luminosity radio-quiet quasar PDS 456, the most
luminous known AGN at $z<0.3$. Repeated X-ray flaring is found in
PDS 456, over the duration of the 340 ksec long \sax\ observation. 
The X-ray flux doubles in just 30 ksec, whilst the total
energy output of the flaring events is as high as $10^{51}$~erg. Under the
assumption
of isotropic emission at the Eddington limit, this implies that the
size of the X-ray emitting region in PDS 456 is less than 3
Schwarzschild radii, for a $10^9$M$_{\odot}$ black hole. From
the rates of change of luminosity observed during the X-ray flares,
we calculate lower limits for the radiative efficiency limit between
0.06 and 0.41, implying that accretion onto 
a Kerr black hole is likely in PDS 456.
We suggest that the rapid
variability is from X-ray flares produced through magnetic
reconnection above the disc and calculate
that the energetics and timescale of the flares are plausible if the
quasar is accreting near to the maximum Eddington rate. 
A similar mechanism may account for the extreme rapid
X-ray variability observed in many Narrow Line Seyfert
1s. In the case of PDS 456, we show that the X-ray flaring
could be reproduced through a self-induced cascade of $\sim1000$
individual flares over a timescale of the order 1 day.

\end{abstract}


\section{Introduction}

PDS 456 is a luminous, butlow redshift ($z=0.184$) radio-quiet quasar
identified in 1997 (Torres \et 1997). The optical and
infra-red spectra
(Simpson et al. 1999)
show broad Balmer and Paschen lines (e.g. H$\beta$ FWHM 3000 km~s$^{-1}$),
strong Fe \textsc{ii}, a hard (de-reddened)
optical continuum ($f_{\nu} \propto \nu^{-0.1\pm0.1}$), and one of the
strongest `big blue bumps' of any AGN (Simpson \et 1999, Reeves et
al. 2000). It is also radio-quiet (F$_{5GHz}=8$mJy;
Reeves et al. 2000), and is presumably not jet dominated or strongly
beamed. PDS~456 has a de-reddened, absolute blue magnitude of
M$_{B}\approx -27$ (Simpson et al. 1999),
making it as luminous as the radio-loud quasar
3C~273 ($z=0.158$, M$_{B}\approx -26$).
Indeed PDS 456 is the most luminous known AGN in the local Universe
(z\ $<0.3$), its luminosity being more typical of quasars at z=2-3, at the
peak of the quasar luminosity function. 

PDS 456 was first detected as the X-ray source
RXS~J172819.3-141600 in the ROSAT All Sky Survey (Voges \et
1999). Subsequent \asca\ and \xte\ observations of PDS 456
showed that it was highly X-ray variable (see Reeves et al. 2000).
In particular, during the \xte\ observation,
an X-ray flare occurred
with a doubling time of just 15 ksec, implying that the X-ray
emitting region was extremely compact, less than 2 Schwarzschild radii
(or $2R_S$) in size. Such rapid
variability is very unusual for luminous quasars, as the
variability timescale is thought to increase with luminosity,
and black hole mass (e.g. Turner \et 1999). One possibility is that
the accretion rate is unusually 
high in PDS 456, perhaps close to Eddington. An
analogy might then be drawn with the extreme events observed in several 
Narrow Line Seyfert 1 galaxies (e.g. Boller, Brandt \& Fink 1996,
Leighly \et 1999), thought to have smaller black hole masses
($10^6$M$_{\odot}$ - $10^7$M$_{\odot}$), accreting near to the
Eddington rate.
We report here on X-ray observations of PDS 456, conducted
with \sax\ and \xmm\ in February and March 2001. 
%This paper concentrates on
%the X-ray variability, observed primarily with the \sax\
%satellite. 
The prime motivation was to study the
extra-ordinary variability of PDS 456 with an imaging X-ray
telescope, thus negating the possibility of source contamination
which may occur within the field of view of a non-imaging instrument.
%A detailed account of the spectral analysis from the \xmm\
%observations will be published separately (Reeves et al. 2002).
%Values of $ H_0 = 50 $~km\,s$^{-1}$\,Mpc$^{-1}$ and $ q_0 = 0.5 $ have been
%assumed throughout and all fit parameters are given in the quasar
%rest-frame.
%Note that errors in this paper are quoted at
%the 90\% confidence level.

\begin{figure}
\begin{center}
\rotatebox{-90}{\includegraphics[width=6.5cm]{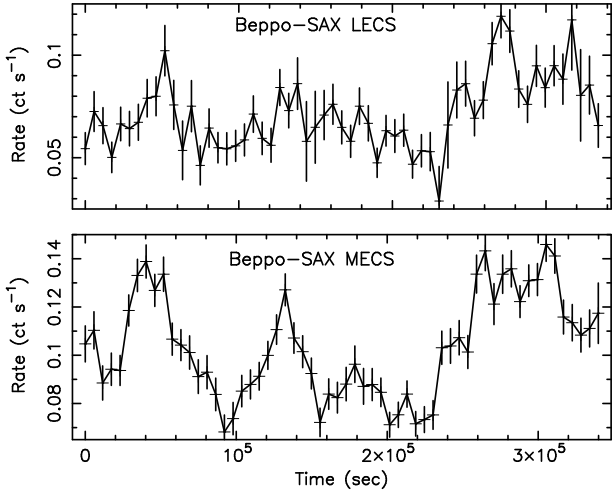}}
\end{center}
\caption{Background subtracted \sax\ lightcurves for the (a) LECS (0.3-2
keV) and (b) MECS (1-10 keV), binned into orbital bins (96 minutes).
Repeated X-ray flaring is seen throughout the observation.
The largest changes are by a factor of x1.9 in the MECS in 35 ksec
and a factor of x4.1 in the LECS in 40 ksec, after 230 ksec.
%Further significant flux increases of x1.7
%and x1.9 are seen in the MECS, at 20 ksec and 100 ksec, over rise times of
%30 and 35 ksec respectively.
The variations imply that for PDS 456,
with a $10^9$M$_{\odot}$ black hole, the
X-ray emitting region is no larger than 3 Schwarzschild radii ($3R_S$).}
\end{figure}

\section{The X-ray Observations}

PDS 456 was observed by \sax\ between 26th February 2001 and 3rd March 2001,
with a total duration of 345~ksec. Lightcurves were
extracted from circular regions of 4\arcmin\ and 6\arcmin\ radius
around PDS 456, for the MECS (Medium Energy Concentrator Spectrometer;
Boella \et 1997) and the LECS (Low Energy Concentrator Spectrometer; Parmar
\et 1997) detectors respectively. Background lightcurves
were taken from circular regions, offset from the source. 
The average net source count rates obtained were
$(1.045\pm0.001)\times10^{-1}$~ct~s$^{-1}$
and $(7.25\pm0.12)\times10^{-2}$ct~s$^{-1}$ for the MECS and LECS
respectively, whilst the background rates were
%$(6.67\pm0.21)\times10^{-3}$ct~s$^{-1}$ and
%$(5.9\pm0.6)\times10^{-3}$~ct~s$^{-1}$, 
less than 10\% of the source count rate.
\xmm\ also observed PDS 456 on 26th February 2001, with a duration of
40 ksec. The start of the \xmm\ observation was coincident with the
onset of the \sax\ exposure. Here, we show only the
timing data from the EPIC-pn detector (Struder \et 2001). 
A full description of the \xmm\ data and
spectroscopic observations will
be presented in a subsequent paper (Reeves \et 2002, in preparation).
Values of $ H_0 = 50 $~km\,s$^{-1}$\,Mpc$^{-1}$ and $ q_0 = 0.5 $ have been
assumed throughout and errors are quoted at
the 90\% confidence level.

\section{The X-ray Variability of PDS 456}
 
The background subtracted \sax\ lightcurves are plotted in Figure 1, 
grouped into orbital length bins (96 minutes), and extracted over the 
energy bands 0.3-2 keV and 1-10 keV, for the LECS and MECS respectively. 
Large changes in flux are seen throughout the \sax\
observation. The most extreme events are the flares seen toward the
end of the observation at 230 ksec, where the MECS count rate
increases by a factor of $\times$1.9 over 35 ksec, whilst the LECS flux
increases by $\times$4.1 over 40 ksec. 
%Pronounced events are also seen earlier
%in the MECS observation, an increase by a factor $\times$1.7 in 30 ksec after
%20 ksec and an increase of $\times$1.9 in 35 ksec after 100 ksec, whilst the
%LECS curve rises by a factor $\times$2 in 35 ksec after 20 ksec.

The \xmm\ EPIC-pn lightcurve is plotted in Figure 2 (upper
panel), extracted from 0.3-10 keV. The increase in flux 
corresponds to the initial rise of the \sax\ lightcurves over its
first 40 ksec of observation, where these two observations were
concurrent. There is no variability over timescales as short as
$\sim$~1~ksec, consistent with a large ($10^{9}$M$_{\odot}$)
black hole mass in this quasar. Using the \xmm\ data, we also 
searched for evidence of spectral variability in PDS 456. A `softness ratio'
was defined, taking the ratio of the pn count rate over the 0.3-1.0
keV and 2-10 keV bands; the time-averaged value was then renormalized to 1.
%Note that in the \xmm\ EPIC spectrum of PDS 456 (Reeves \et
%2002), a strong soft X-ray excess dominates the flux in the 0.3-1.0 keV
%band, which may be associated the high energy Wien tail of the so-called
%`Big Blue Bump' (e.g. Malkan \& Sargent 1982, Elvis \et 1994), whilst
%power-law emission dominates the 2-10 keV band. 
A plot of softness ratio
against time is
shown in Figure 2 (lower panel). The spectrum becomes softer (by a
factor $\sim$20\%) when the count rate is higher during the course of the
flare in the pn lightcurve. This is also illustrated in Figure 3, 
which plots the
softness ratio as a function of total pn count rate, the correlation is
significant at $>99.99$~\% confidence using a Spearman-Rank test. 
%Performing the
%same analysis on the \sax\ data gives a similar correlation, albeit with
%lower signal to noise.

We also constructed cross correlation functions in order
to search for any soft to hard time lags using both the \sax\ and
\xmm\ data.  No delays were found, the upper-limit 
was $<1$~ksec. It appears that both
the soft (0.3-1.0) keV and hard (2-10 keV) bands vary coherently on
timescales much shorter than that of the overall flare duration. For simple
reprocessing models, where the thermal disc emission is Compton
up-scattered to reproduce the hard X-ray power-law (e.g. Czerny \& Elvis
1987), this implies that the size of the reprocessing region is
$<10^{13}$~cm or
$<0.1R_S$ (for a Thomson depth of $\tau\sim1$). The spectral
softening appears consistent with observations of some AGN,
where the X-ray spectra are generally softer at higher fluxes (e.g. Vaughan
\& Edelson 2001).

\section{The Size and Efficiency of the X-ray Emission in PDS 456}

Using light-crossing arguments, and assuming that relativistic beaming
is unimportant, one can calculate the overall size of the X-ray emitting
region in PDS 456 from the expression 
$l=ct/(1+ \tau)$, where t is the rise-time of
the flares and $\tau$ is the Thomson depth of the X-ray emitting
region. Past multi-wavelength studies have shown that PDS 456
has a total bolometric luminosity of $10^{47}$~erg~s$^{-1}$, peaking in
the optical-UV band (Simpson \et 1999, Reeves \et 2000). With the
assumption of isotropic emission (noting that
PDS 456 is a radio-quiet, non-blazar AGN), 
this then requires a $10^9$M$_{\odot}$ black
hole accreting at the Eddington rate. As the X-ray flux from PDS 456
can double on timescales of $\sim30$~ksec, then the X-ray
emitting region has a maximum size of $l<10^{15}$~cm, or
$<3R_S$ for a $10^9$M$_{\odot}$ black hole, consistent with
the previous \xte\ measurement in 1998 (Reeves \et 2000). If one relaxes
the assumption that PDS 456 is accreting at the Eddington rate, and
thus the black hole mass is $>10^9$M$_{\odot}$, or if the Thomson
depth of the region is large ($\tau>1$), 
then the restriction on the radius of the emitting region is even more severe
(i.e. $l<<3R_S$). Thus the extreme variability 
suggests that the accretion rate in PDS 456 is indeed very high, with
the X-ray emission originating from a very compact region (of $<3R_S$),
presumably close to the event horizon of the putative $10^9$M$_{\odot}$
black hole.

\begin{figure}
\begin{center}
\rotatebox{-90}{\includegraphics[width=5.5cm]{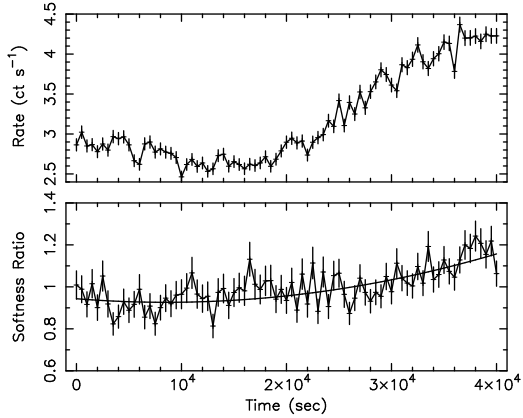}}
\end{center}
\caption{\xmm\ EPIC-PN lightcurve of PDS 456, concurrent with the
first 40 ksec of \sax\ observation (upper panel).
The observations are consistent with the \sax\ lightcurves, an increase in
flux
is seen after 20 ksec in \xmm\, but there are no rapid changes
over shorter timescales. The lower panel shows the variation in
softness ratio (0.3-1.0/2-10 keV count rate) versus time, which shows the
source softening towards the end of the observation.}
\end{figure}

\begin{figure}
\begin{center}
\rotatebox{-90}{\includegraphics[width=5.5cm]{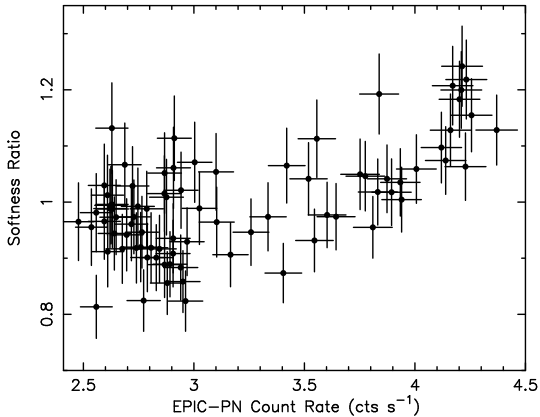}}
\end{center}
\caption{Softness ratio verus total PN count rate. A significant
correlation is seen, showing that the spectrum of PDS 456 softens with
increasing flux.}
\end{figure}

It is possible to calculate the efficiency of converting
rest mass into energy from the flaring events. 
Firstly we estimate the total X-ray luminosity
from the \xmm\ spectral fits to PDS 456. The \xmm\ EPIC MOS
and pn spectrum of PDS 456 is shown in Figure 4; extrapolating a  
simple power-law fit to the 2-10 keV spectrum shows a 
clear excess of counts below 1~keV in both detectors.  Fitting the
continuum with a hard power-law (of $\Gamma=1.98\pm0.02$) and a single
temperature blackbody component to model the soft excess (with
$kT=101\pm5$~eV), then the time-averaged 0.3-10 keV luminosity obtained
is $1.10\times10^{46}$~erg~s$^{-1}$, compared to the bolometic
output of $10^{47}$~erg~s$^{-1}$ (Simpson \et 1999, Reeves \et 2000). 
Note that slightly higher values for the luminosity are obtained if
one uses either a Comptonised blackbody or a disk blackbody spectrum. 

The time-averaged count rate
in the EPIC-pn detector is $3.33\pm0.01$~ct~s$^{-1}$, hence one can
calculate a constant factor for the conversion of count rate to luminosity of
$f_{pn}=(3.30\pm0.01)\times10^{45}$~erg~ct$^{-1}$.  
From a linear fit to the increase in the \xmm\ lightcurve after 20
ksec (Figure 2),
we find a change in PN count rate of
$1.64\pm0.13$~ct~s$^{-1}$, over $18/(1+z)$~ksec (quasar rest
frame), a factor of $\times1.62$
increase. Thus the corresponding rate of change in
luminosity of PDS 456 is
$\Delta L/\Delta t=(3.56\pm0.27)\times10^{41}$~erg~s$^{-2}$. Assuming
photon diffusion through a spherical mass of accreting matter in which
the opacity is dominated by Thomson scattering, the observed
change in luminosity implies a {\it minimum} efficiency of 
$\eta>(\Delta L/\Delta
t)/(2\times10^{42})$ (Fabian 1979, Guilbert, Fabian \& Rees 1983).  
The derived efficiency, from this expression, of $\eta>0.18\pm0.02$,
exceeds the theoretical maximum for a Schwarzschild 
black hole, but is consistent with the limits for a
Kerr metric (Thorne 1974).

Constraints can also be placed on $\eta$ from the
\sax\ LECS lightcurve. A 0.3-2.0 keV luminosity of
$9.65\times10^{45}$~erg~s$^{-1}$ was derived from a
power-law plus blackbody fit to the
(high signal to noise) \xmm\ spectrum. The mean LECS count rate,
over the portion of the lightcurve simultaneous with \xmm\,
was $(6.5\pm0.2)\times10^{-2}$~ct~s$^{-1}$. Hence the constant
factor for converting between luminosity and \sax\ LECS count
rate is $f_{LECS}=(1.48\pm0.07)\times10^{47}$~erg~ct$^{-1}$. 
The two fastest events observed here correspond to 
an increase in count rate of $(5.42\pm1.53)\times10^{-2}$~ct~s$^{-1}$
in $11500/(1+z)$~sec after 230 ksec, and an increase of 
$(4.98\pm0.95)\times10^{-2}$~ct~s$^{-1}$ in $17280/(1+z)$~sec (after 253
ksec), corresponding to $\Delta L/\Delta
t=(8.3\pm2.3)\times10^{41}$~erg~s$^{-2}$ and $\Delta L/\Delta
t=(5.1\pm0.9)\times10^{41}$~erg~s$^{-2}$ respectively. 
The implied efficiency
factors are then $\eta>0.41\pm0.11$ and $\eta>0.26\pm0.05$, 
near the maximum permitted value of $\eta\sim0.3$ for extraction of 
energy around a Kerr black hole. 

As discussed by Brandt \et (1999), there are several caveats
to note about the standard definition of the efficiency limit (Fabian
\et 1999, Guilbert \et 1983). The limit can be invalid if 
the X-ray emission is relativistically boosted, as can occur in a blazar
like AGN, although we note here that PDS 456 is radio-quiet (the
radio-loudness of PDS 456, defined as the ratio of 5 GHz to B band flux, is
$F_{6cm}/F_{B}=0.18$) - hence
any X-ray emission associated 
with a radio-jet is likely to be weak or absent.  
Mild relativistic boosting may be possible through coronal flare
emission, in this case velocities of $\beta=v/c=0.3$ can be achieved
(e.g. Beloborodov 1999). However in PDS 456 the relatively steep
($\Gamma=2.0$) time-averaged 2-10 keV photon index, and the 
spectral steepening observed during a flare (Figures 2 and 3), 
makes this possibility unlikely.

\begin{figure}
\begin{center}
\rotatebox{-90}{\includegraphics[width=6cm]{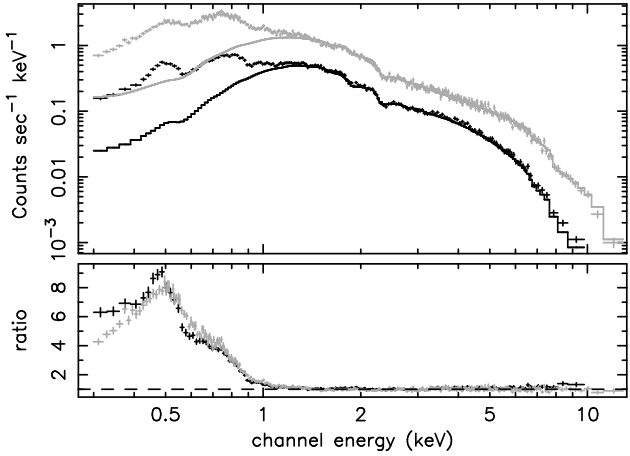}}
\end{center}
\caption{The \xmm\ EPIC MOS (black) and pn (grey) spectrum of PDS
456. A power-law of photon index $\Gamma=1.98\pm0.02$
has been fitted to the 2-10 keV spectrum. Extraploation of this
power-law model to lower energies (0.3 keV) results in a clear soft
X-ray excess, which can be modelled by either blackbody or
comptonised blackbody emission. The total X-ray luminosity measured
from 0.3-10 keV is then $1.1\times10^{46}$~erg~s$^{-1}$.}
\end{figure}

The derived efficiency also assumes that the radiation is emitted at
the centre of a spherical region; the limit can become invalid
if the X-ray emission occurs in the outer layers of a sphere of high
($\tau>>1$) optical depth, as the flare rise-time can then become
unusually short (see the discussion in Brandt \et 1999, Appendix
A). Such a high value for $\tau$ seems unlikely in PDS 456, as from
light-crossing arguements this
would require a very compact region of $l<<3R_{S}$  
for the X-ray emission. Nonetheless, in order to adopt a
more conservative approach in estimating $\eta$, we calculated the total  
{\it integrated}
increase in X-ray emission during the course of a flare, rather than
using the rise-time of the event. The integrated energy emitted
during the largest flare during the \sax\ observation (after 230 ksec)
was $(9.1\pm0.6)\times10^{50}$~erg, over a duration of
$103.6/(1+z)$~ksec. Hence the rate of change of X-ray luminosity 
over the whole flare is $\Delta L/\Delta
t=(1.19\pm0.08)\times10^{41}$~erg~s$^{-2}$ and the derived efficiency
limit is then $\eta>0.060\pm0.004$, close to the maximum possible value
for accretion onto a Schwarzschild black hole.

\section{Discussion}

The tight limits placed on the timescale and size of the
X-ray variations in PDS~456 allow us to place constraints on its
physical origin. The rapid rise times and large amplitudes 
of the X-ray variations suggest that these are 
coherent events from the innermost regions ($R_{\rm in} \sim 3 R_S$) 
of the accretion disc.  Moreover the high value of the 
radiative efficiency, measured between $\eta>0.06$
(conservative) and $\eta>0.41$ (optimal),  
is indicative of accretion onto a Kerr black 
hole. The timescale of the flares ($t_{\rm flare} \sim 100$ ksec)
is shorter than the dynamical time scale ($t_{\rm dyn}$) 
of the disc at 3$R_S$  (here $t_{\rm dyn} \sim 350$ ksec at
3$R_S$, for a pseudo-Newtonian potential, Abramowicz 1996).
We suggest that X-ray flares produced via magnetic 
reconnection events in the corona of the inner disc are 
a plausible mechanism to explain the observed variations over such a
short (sub-orbital) time scale. 

Magnetic reconnection events are likely to arise as a direct result of
buoyant magnetic flux tubes emerging randomly from the disc surface.
Reconnection then takes place in regions of the
corona in which oppositely directed magnetic field lines
come in close contact. 
Di Matteo (1998) examined coronal heating  
and flare production via Petschek reconnection events 
in AGN. The discussion presented here is based on many of the
results of that analysis. 
Flare events in many AGN can be accounted for by the onset of an 
ion-acoustic instability associated with slow MHD
shocks and Petschek reconnection in the accretion disc corona. 
The fundamental energy source driving this 
process is the orbital energy of the inner disc, 
as shear stresses in the turbulent
disc can rapidly amplify any seed fields. The resultant
flux tubes are subject to a buoyancy (Parker) instability
and emerge from the disc into the magnetically-structured
corona. In this way a fraction of the accretion energy is
converted into coronal magnetic energy. 
The emergence rate of magnetic energy into the disc corona 
should therefore be proportional to the the kinetic energy 
flux through the inner disc
$\dot{E}_{\rm disc} = {\partial}_{\rm t}(2\pi R \Sigma
v_\phi^2) \sim \dot{M} v^2_\phi \sim \dot{M} c^2 / \epsilon$ 
in steady state at $R = \epsilon R_{\rm S}$, where $\Sigma$ is the 
disc surface density, $v_\phi$ is the azimuthal velocity and  
$\dot{M}$ is the steady state accretion rate.
Assuming a $10^9 M_\odot$ black hole accreting at the Eddington rate,  
we have $\dot{E}_{\rm disc}(3R_{\rm S}) \sim 10^{47}$ 
erg s$^{-1}$ with $\epsilon = 3$ and from observation the total output
of the flares is
%$E_{\rm flare} \sim \dot{E}_{\rm disc} t_{\rm rec} \sim 10^{51}$
$E_{\rm flare} \sim 10^{51}$~erg, over the lifetime (recurrence time) 
of the flares of $t_{\rm rec}\sim 1$ day. 
A relatively high efficiency ($\ga 0.1$)
for the conversion of accretion energy to X-ray luminosity 
is required to power the flares in PDS~456, and is 
another indication that the massive black hole in this object  
is accreting at a rate close to the Eddington limit.

%\begin{equation} \label{buoy}
%\frac{B_{\rm c}^2}{8\pi} \sim n_{\rm disc} k T_{\rm disc}
%\end{equation}

The buoyancy condition for magnetic flux tubes within the accretion
disc is that the magnetic pressure should exceed
the local gas pressure (e.g. Coroniti 1981), the critical condition
being $\frac{B_{\rm c}^2}{8\pi} \sim n_{\rm disc} k T_{\rm disc}$, 
where $B_{\rm c}$ is the local magnetic field strength, $n_{\rm disc}$ is the 
particle density within the disc and $T_{\rm disc}$ is the disc
temperature. The structure of the accretion flow close to the last stable
orbit is unknown, however estimates 
can be made by assuming that accretion takes place via a thin disc.
Using a steady-state
Shakura-Sunyaev disc with a viscosity parameter  
$\alpha \sim 0.1$ yields the estimates
$n_{\rm disc} \sim 10^{17}$ cm$^{-3}$ and $T_{\rm disc} \sim 10^5$ K at 
$r \sim 3R_S$. Applying these estimates to the magnetic pressure
condition above, leads to a value for the field
strength of a buoyant flux tube within the inner regions
of the disc of $B_c \sim 5000$ G. 

The observations provide two constraints on the 
density of the X-ray emitting region. These arise from the  
constraints on the fitted column density and the ionisation parameter
of the highly ionised absorber in PDS 456. This is apparent 
in the hard X-ray spectrum of PDS 456 above 7 keV, in the form 
of deep K-shell edges of highly ionised iron (Fe \textsc{xxv} and 
Fe \textsc{xxvi}), 
present both in the current \xmm\ EPIC spectrum (Reeves \et 2002,
in preparation) and in earlier \asca\ and \xte\ observations 
(Reeves \et 2000). From the \xmm\ observations, 
the X-ray spectrum of PDS 456 is well fit with a column density 
$n_{\rm H} \sim n R_{\rm in}$ in the range $9\times 10^{23}$ 
- $5\times 10^{24}$ cm$^{-2}$ implying $9\times 10^8$ cm$^{-3}$
$< n <$ $5\times 10^9$ cm$^{-3}$. On the other hand, the ionisation parameter
$U \simeq L_{\rm x}/n R^2_{\rm in}$ lies in the range $3\times 10^4$ -
$2.5\times 10^5$, where $L_{\rm x} \sim 10^{45}$ erg s$^{-1}$ is the
hard X-ray ionising flux leading to the limits $4 \times 10^9$ $< n <$ 
$3 \times 10^{10}$ cm$^{-3}$. 
If we associate this density with the hot post shock gas 
we obtain the estimate $n_s \sim 5 \times 10^9$ cm$^{-3}$.
The evolution of the buoyant flux tubes provides a natural 
explanation for this very low value of the density of the X-ray 
emitting region. As the flux tubes emerge from the disc and rise
into the corona, gravitational downflow of plasma causes the density
within the flux tube to decrease to the point at which reconnection
can take place efficiently. The density estimate above 
implies that the buoyant flux tubes rise far into the corona
before reconnection occurs. An estimate of the height at which
reconnection occurs in terms of the pressure scale height of the
disc ($H$) can be obtained from (Di Matteo 1998)
\begin{equation}
h_{\rm flare} \sim \biggl ( \frac{n_{\rm disc}}{n_s} \biggr )^{1/4} H
\sim 50 H.
\end{equation}
In the case of a Shakura-Sunyaev accretion disc $H/R \la 0.1$ 
implying that $h_{\rm flare} \ga R_{\rm in}$. 
%Given that the
%estimates of $ n_{\rm disc}$ and $n_s$ are upper and lower
%limits respectively we will assume $h_{\rm flare} \sim R_{\rm in}$.
Hence the total magnetic energy stored in the  
corona of the inner disc should be of the order 
$B_c^2 R_{\rm in}^3 / 8 \pi \sim 10^{51}$ erg. 
This large reservoir of stored
accretion energy is enough to explain the large amplitude 
X-ray variations seen in PDS~456, and highlights the
efficiency with which accretion energy must be converted 
to magnetic energy to explain these flaring events.

%Thus far we have argued that the energetics of the accretion
%process in PDS~456 can explain the observed variations
%in the X-ray light curve as long as the black hole mass
%$M_{\rm BH} \sim 10^9$  $M_\odot$, $\dot{M} \simeq \dot{M}_{\rm Edd}$ 
%and the kinetic energy of the accretion flow can be
%efficiently converted into magnetic energy within
%the disc corona. 
We now consider the rise time 
of the events in the light curves ($t_{\rm rise} \ga
30$ ksec) in the light of the above model.
An attractive feature of the Petschek model is the short time scale on which
reconnection can take place. This can be of the order of a few Alfv\'{e}n 
times $\tau_{\rm A} \sim l / v_{\rm A} \sim R_{\rm in} 
(4\pi n_s m_{\rm p})^{1/2}/ B_c$ 
where $l$ is the characteristic length scale associated with changes of 
the magnetic field (here $l \sim R_{\rm in}$), 
$v_{\rm A}$ is the local Alfv\'{e}n speed
and $m_p$ is the proton mass. The estimates above yield 
$\tau_{\rm A} \sim 50$ ksec, encouragingly close to the
observed rise time of the flares. 
The time scale $\tau_{\rm A}$ provides a constraint on the 
density of the accretion disc and post 
shock gas. An accretion disc density significantly lower than
$n_{\rm disc} \sim 10^{17}$ cm$^{-3}$ or a shock
density  much higher than $n_s \sim 10^{10}$ cm$^{-3}$ would
increase $\tau_{\rm A}$ to a level inconsistent with the
observations. 
%We therefore conclude that the density estimates
%above are good to order of magnitude and that Petschek 
%reconnection events can explain the timescale of the X-ray 
%variability of PDS~456. 

The energy production rate for an individual flare can be estimated 
from (Di Matteo 1998):- 
\begin{equation}
\dot{E}_{\rm flare} \sim \frac{n_s L l (k T_{\rm flare})^{3/2}}
{m_e^{1/2}} \sim \frac{n_s R_{\rm in}^2  
(k T_{\rm flare})^{3/2}} {m_e^{1/2}}
\end{equation}
where $L$ is the length of the slow shock region, 
$T_{\rm flare}$ is the X-ray temperature and $m_e$ is the electron mass.
The dimension $L$ is the essentially the length of region
of oppositely-directed magnetic field lines which we constrain
to be of the order $L \sim R_{\rm in}$. 
Adopting $T_{\rm flare} \sim 10^9$ K we obtain $\dot{E}_{\rm flare}
\la 10^{43}$ erg s$^{-1}$. This suggests that the large scale
X-ray variations in the light curve of PDS~456 involve $\ga$ 1000 
or so individual flares. These flares could not produce the
large amplitude variations observed in PDS~456 if they were
incoherent. 

However it is possible that the magnetic 
structure within the disc corona could reach
a self-organized, critical state in which the reconnection 
and flaring of one flux tube could prompt similar flares
in its neighbors, allowing a coherent cascade of flares
to develop (Leighly \& O'Brien 1997). 
The timescale on which the global cascade could take place 
can be constrained from the Alfv\'{e}n time in the region 
involved, which is consistent with observed timescale of the flaring
events in the light curve.

This leads to a suggestion of why it is that 
PDS~456 releases such a large fraction of its accretion
energy in the form of large-amplitude, coherent X-ray
variations. A flare cascade along the lines of that suggested 
above would require that magnetic energy was stored in the 
disc corona until some triggering criterion was reached,
promoting the first flare event. The energy involved in 
such a cascade is likely to be governed by the time scale on
which magnetic energy is pumped into the disc corona
($t_{\rm mag}$) and the time scale on which the triggering
criterion is satisfied ($t_{\rm trigger}$). Two possible cases
emerge: (a) $t_{\rm mag} \gg t_{\rm trigger}$ in which case
the flares will be incoherent and the resultant X-ray 
variability small-amplitude  and (b) $t_{\rm mag} \la 
t_{\rm trigger}$ producing the possibility of large-amplitude
coherent flare cascades. Unfortunately
little is known about $t_{\rm trigger}$, however $t_{\rm mag}$
must be related to $\dot{E}_{\rm disc}$ via 
$t_{\rm mag} \sim E_{\rm c}/\dot{E}_{\rm disc}$ where 
$E_{\rm c}$ is a measure of the minimum magnetic energy 
required to be stored in the disc corona before it can
reach a critical state allowing a flare cascade. Hence
$t_{\rm mag} \propto E_{\rm c}/\dot{M}
\propto E_{\rm c}/\dot{m}M_{\rm BH}$, where $\dot{m}$ is the 
accretion rate in Eddington units. Hence systems in which a 
high mass black hole accretes at the Eddington limit should
be more able to release stored magnetic energy in the
form of a coherent flare cascade, providing at least a tentative 
explanation of why the X-ray variability of PDS~456 is so extreme. 
The similarity of the X-ray behavior
of PDS~456 and the Narrow Line Seyfert 1s (see Leighly \et 1999) 
suggests that $\dot{m}$ is the critical
factor in determining the scale of the X-ray variability in AGN.
The energy of X-ray flares in some NLS1s can be as high as $\sim 10^{48}$
erg (e.g. IRAS 13324-3809, Boller \et 1997), 
a factor of $\sim10^3$ less than the flares
observed in the light curve of PDS~456, in good
agreement with the ratio of black hole masses in these
systems. If we use these observations to make the connection $E_{\rm c}
\propto M_{\rm BH}$, then $t_{\rm mag} \sim 1/\dot{m}$. Such a 
relation would suggest that any system accreting close to the
Eddington rate would be likely to show strong X-ray variability.

\section{Conclusions}

Recent \sax\ and \xmm\ observations have shown that the luminous
quasar PDS 456 exhibits rapid X-ray variability on  
timescales of $\sim30$~ksec, with a total energy output of 
$10^{51}$~erg~s$^{-1}$ for the flaring events. 
This limits the size of the 
X-ray emitting region to $<3$ Schwarzschild radii for a 
$10^{9}$M$_{\odot}$ black hole. The energetics of the 
of the accretion disc in PDS~456 can power its extreme
X-ray variability if the black hole is massive
($\ga 10^9$ $M_\odot$) and is accreting close to the
Eddington rate. Coronal magnetic
flare events can explain the X-ray variability as long as
the disc is able convert accretion energy into 
coronal magnetic energy efficiently, and that this energy 
can be released in the form of a self-induced cascade of $\ga$ 1000 
individual flare events on a timescale of the order 1 day.

%\bsp
\label{lastpage}
\end{document}